# Robust multicolor single photon emission from point defects in hexagonal boron nitride

*Toan Trong Tran,[1] Christopher ElBadawi,[1] Daniel Totonjian,[1] Charlene J Lobo,[1] Gabriele Grosso,[2] Hyowon Moon,[2] Dirk R. Englund,[2] Michael J. Ford,[1] Igor Aharonovich [1]\* and Milos Toth[1]^*

[1] School of Mathematical and Physical Sciences, University of Technology Sydney, Ultimo, NSW, 2007, Australia

[2] Department of Electrical Engineering and Computer Science, Massachusetts Institute of Technology, Cambridge, Massachusetts, 02139, United States

\*igor.aharonovich@uts.edu.au
^milos.toth@uts.edu.au

**Abstract**

Hexagonal boron nitride (hBN) is an emerging two dimensional material for quantum photonics owing to its large bandgap and hyperbolic properties. Here we report a broad range of multicolor room temperature single photon emissions across the visible and the near infrared spectral ranges from point defects in hBN multilayers. We show that the emitters can be categorized into two general groups, but most likely possess similar crystallographic structure. We further show two approaches for engineering of the emitters using either electron beam irradiation or annealing, and characterize their photophysical properties. The emitters exhibit narrow line widths of sub 10 nm at room temperature, and a short excited state lifetime with high brightness. Remarkably, the emitters are extremely robust and withstand aggressive annealing treatments in oxidizing and reducing environments. Our results constitute the first step towards deterministic engineering of single emitters in 2D materials and hold great promise for the use of defects in boron nitride as sources for quantum information processing and nanophotonics.

Hexagonal boron nitride (hBN) is a van der Waals material that has recently emerged as a fascinating platform for room temperature quantum photonics due to the discovery of room temperature quantum emitters,[1] realization of sub-diffraction focusing and guiding,[2,3] super-resolution imaging,[4] and tunable phonon polariton propagation.[5,6] While the optical properties of bulk hBN have been studied thoroughly[7-9], detailed photo-physical properties of its two dimensional counterpart are scarce. In particular, the luminescent properties of hBN under sub-band-gap excitation remain largely unexplored.

In traditional 3D semiconductors, including diamond and silicon carbide, color centers have similar spectral properties in both bulk and nanostructured forms. This may be advantageous when controlled engineering of defects is required. Both materials can be doped during growth, resulting in deterministic formation of luminescent centers. However, in van der Walls crystals, the electronic and optical properties of two dimensional (single or few monolayer) flakes are significantly different from their bulk counterparts.[10] This often results in fascinating phenomena such as spin valley splitting[11] or strong exciton-phonon interactions that can be observed at room temperature,[12] but poses major challenges for engineering and control of single color centers.

In this letter, we report an unprecedented phenomenon – namely, narrowband multicolor single photon emission from a 2D material. While standard quantum dots and known color centers luminesce at a particular wavelength or a narrow spectral range, we show that defects in hBN multilayers can emit over a broad range spanning over 200 nm. We also show that the emitters can be engineered using electron irradiation and withstand various aggressive annealing treatments in reactive gaseous environments, which do not change their spectral properties. Our results pave the way to robust, room temperature quantum photonic devices that employ color centers in hBN as key building blocks.

Figure 1a shows a simplified schematic of the confocal photoluminescence (PL) setup used to characterize hBN. A detailed schematic and a typical confocal map featuring a number of single photon emitters are shown in Figure S1 of the Supporting Information. Unless noted otherwise, the measurements were performed at room temperature using either a 532 nm continuous-wave (CW) laser, or a 510 nm pulsed laser as an excitation source for lifetime measurements. We performed a photoluminescence (PL) survey and collected spectra of various single defect centers in hBN (which was annealed previously in argon at $850^O$ C, as is discussed below). A representative range of room temperature PL spectra is shown in Figure 1(b, c). The emitters have narrow zero phonon lines (ZPLs) at energies in the range of ~1.6 – 2.2 eV (~ 565 – 775 nm).

The emitters can be classified into two general groups based on their ZPL energy and phonon side band (PSB) spectral shapes. Group 1 (Figure 1b) consists of emitters with ZPL energies of 576 nm (2.15 eV), 583 nm (2.13 eV), 602 nm (2.06 eV), 633 nm (1.96 eV), and 652 nm (1.90 eV). Emitters in this group exhibit relatively broad and asymmetric ZPL line shapes with pronounced low energy tails. The spectra also contain pronounced doublet PSBs. Group 2 (Figure 1c) is comprised of emitters at lower energies, with ZPLs centered on 681 nm (1.82 eV), 696 nm (1.78 eV), 714 nm (1.74 eV), and 762 nm (1.63 eV). Notably, these emitters have narrower, more symmetric ZPLs, with phonon sidebands that are weak compared to Group 1. A survey of ~ 40 emitters (Figure S2(a-e)) in the sample annealed at 850°C revealed a relative abundance of ~70% and 30% of emitters in Groups 1 and 2, respectively. Figure 1d is a histogram showing the ZPL energies of emitters classified into the two groups.

A Hanbury Brown and Twiss (HBT) setup was used to verify single photon emission from these defects. Figure 1e shows second-order autocorrelation functions ($g^{(2)}(\tau)$) recorded from representative emitters selected from each group, with ZPLs centered on 633 nm (Group 1 emitter), and 714 nm (Group 2 emitter). Both curves show that $g^{(2)}(0) < 0.5$, proving unambiguously that the defects are point defects that act as single photon emitters (an additional $g^{(2)}(\tau)$ from each group is shown in Figure S2(f,g)). The data were not background-corrected[13], fit using a three level model, and offset vertically for clarity.

It is important to note that emitters in both groups exhibit a similar energy gap difference of 160 ± 5 meV between the ZPL and the PSB (Figure 1f). These values indicate that the associated coupled localized vibrations (phonon modes) are very similar, and the defects responsible for all spectra have similar crystallographic structure.[14, 15] Hence, the two groups likely correspond to two similar defects that reside in different local dielectric environments. The variation in ZPL position within each group may be contributed to by variations in local strain and dielectric environment, as is discussed below. We note that the spectra of emitters in Group 1 are very similar to those of a color center that was previously ascribed to the $N_BV_N$ defect in hBN.[1]

We now turn to fabrication methodologies used to engineer the emitters. We developed two different processes based on annealing and electron beam irradiation, illustrated schematically in Figure 2a. Both methods can be used to create the studied defects in hBN. The annealing method was optimized by varying the annealing temperature of as-grown flakes in an inert environment. Each annealing treatment was performed for 30 minutes under 1 Torr of argon. Figure 2a shows that the number of stable color centers found by confocal PL increases with annealing temperature, indicating that defect diffusion and lattice relaxation occur in the flakes. This behavior is similar to the well-studied nitrogen vacancy center in diamond.[16, 17]

The second process involves electron beam irradiation performed using a scanning electron microscope. The as-grown flakes were first deposited on a silicon substrate and pre-characterized by confocal PL mapping and spectroscopy. Then, particular sample regions were irradiated by a 15 keV, 1.4 nA electron beam for one hour in a low vacuum, $H_2O$ vapor environment[18] (the $H_2O$ prevents electron beam deposition of carbon[19] that is luminescent and modifies PL spectra). A detailed description of the irradiation experiments is provided in the Experimental Methods section. The pre-characterized sample regions were then re-measured using the confocal PL microscope. Figure 2(c,d) shows photoluminescence spectra recorded before (black curve) and after electron irradiation (red curve) from two sample regions. Luminescent defects were created by the electron beam in each case. Importantly, no annealing was performed before or after the electron beam irradiation treatments. Our results therefore demonstrate two distinct robust methodologies for engineering of the emitters in hBN.

Next, we proceeded to study the stability of the emitters in various gaseous environments. These properties are important both from a technological point of view, since the emitters can potentially be used as sensors or quantum light sources in harsh chemical environments, and for understanding their chemical origin, as annealing in different gases can modify defect emission properties[20, 21]. First, we leveraged the defect fabrication study (Figure 2(b)) to create emitters by annealing a sample for half an hour in Ar at 850°C. The sample was then characterized by confocal PL, annealed sequentially at 500°C for one hour each in hydrogen, oxygen and ammonia environments, and re-characterized by PL after each annealing step. Spectra from two stable emitters, representative of Group 1 and 2, are shown in Figure 3 (a, c), respectively. The fluorescence from the emitters remains unmodified even after annealing in both oxidizing and reducing environments. Figure 3 (b, d) shows the corresponding $g^{(2)}(\tau)$ curves recorded for each emitter after the initial argon annealing treatment (black curve) and after the final annealing step performed in an ammonia environment (red curve). The data convincingly prove the robustness of the emitters and the persistence of their quantum nature.

The annealing results provide important insights into the nature of the emitters. First, the luminescent defect is likely to have a vacancy in its crystallographic structure. This is because of the clear increase in formation probability with annealing temperature, a behavior that is very typical of vacancy-related complexes in solids[17, 22]. Second, it is likely that the emitters are charge neutral. If the emitters were negatively charged, annealing in a hydrogen environment would have been expected to modify the charge state and modify or eliminate the emission. This behavior is exemplified by the NV center in diamond which switches from the negative to the neutral charge state upon annealing in hydrogen (and vice versa upon annealing in oxygen).[17, 22] Similarly, many negatively charged emitters in GaN are switched off upon annealing in hydrogen[23]. Positively charged defects are not considered, as to the best of our knowledge positively charged single photon emitters in solids have not been observed. Finally, we believe that the emitters that are stable upon annealing cannot be surface states, as has been observed for some TMDs[24]. This is because surface states are often unstable, and are expected to be modified upon annealing in different reactive environments.

We note, however, that while many of the emitters were absolutely stable and resisted all the annealing treatments, as is illustrated in Figure 3, each annealing step did create some new emitters, and quench some emitters. Both of these effects are demonstrated in Figure S3, and are not surprising, as is discussed in the Supporting Information.

Next, we present a detailed analysis of the photophysical properties of the two emitters whose $g^{(2)}(\tau)$ curves are shown in Figure 1(f), which are representative of emitters in each of the two groups shown in Figure 1(b,c). To obtain the optical transition lifetimes of these emitters, we performed a time-resolved fluorescence measurement using a 510 nm pulsed laser with a 20 MHz repetition rate and 100 ps pulse width. As is seen in Figure 4a, the lifetimes of the two emitters were extracted using single exponential fits, yielding values of 2.9 and 6.7 ns for the centers with ZPLs at 633 nm and 714 nm, respectively (the onsets of the decay curves should be ignored since they correspond to the response of our experimental setup).

Fluorescence saturation behavior was characterized by measuring PL intensity as a function of excitation power. The results are shown in Figure 4b, and the data were fit using the expression: $I = I_\infty P/(P + P_0)$, where $I_\infty$ and $P_{sat}$ are the emission rate and excitation power at saturation,[25] respectively. The resulting emission rates at saturation for the 633 nm and 714 nm emitters are $2.4\times10^6$ and $0.1\times10^6$ counts/s, at $P_{sat}$ = 310 µW and 770 µW, respectively.

To gain further insights into the observed variation in emitter brightness, we measured the autocorrelation functions over longer time scales up to 0.1 seconds. These measurements provide information about the presence of other metastable states with longer decay times.[26-28] Figure 4(c, d) shows long timescale photon antibunching curves for the two color centers. By applying an increasing number of components in the multi-exponential fitting function, we obtained a fitting function with a least number of exponentials and lowest fitting residue $\chi^2$ for each center. Table 1 summarizes the additional decay lifetimes obtained from the fits.

|  | ZPL = 633 nm | ZPL = 714 nm |
| --- | --- | --- |
| $\tau_3$ [µs] | 0.46 | 5.4 |
| $\tau_4$ [µs] | 6.4 | 25 |
| $\tau_5$ [µs] | ---- | 62 |

**Table 1.** Additional metastable states associated with the investigated emitters.

These measurements account for the difference in the overall brightness of the two emitters. The brighter 633 nm defect has fewer metastable states with shorter lifetimes, while the 714 nm defect exhibits multiple additional metastable states with relatively long lifetimes. The differences in lifetimes of both radiative and non-radiative transitions between the two emitters are indicative of local environmental effects such as the presence of neighboring impurities or the proximity of a center to the surface or the edge of a multilayer hBN flake.

To further characterize the photodynamics of these emitters, we obtained photon antibunching curves *versus* excitation power (Figure S4). By fitting the data using the three-level model[29] $g^{(2)}(\tau) = c - ae^{-\tau/\tau_1} + be^{-\tau/\tau_2}$, we obtained the power-dependent emission lifetimes, $\tau_1$, and mestastable state lifetimes, $\tau_2$, as a function of excitation power[30-32]. The corresponding lifetimes $\tau^0_1$ and $\tau^0_2$ were then obtained by extrapolating the data to zero excitation power,[30-32] (Figure S5) yielding the values $\tau^0_1$ = 3.3 and 8.1 ns, and $\tau^0_2$ = 88.6 and 1.2 ns for the 633 nm and 714 nm emitters, respectively. The $\tau^0_1$ values are in good agreement with those obtained by the direct time-resolved fluorescence lifetime measurements discussed earlier.

Finally, we obtained spectra of emitters from each group at cryogenic temperatures. Figure S6 (a, b) shows room temperature and low temperature (14 K) spectra from emitters in Groups 1 and 2 respectively. As expected, the line width is dramatically reduced and approaching 3.87 and 1.17 meV for emitters in Group 1 and Group 2, respectively. The spectra at low temperature confirm the difference in the ZPL shape for the two groups. In particular, the asymmetry of the ZPL in group 1 is maintained at low temperature highlighting a supplementary phonon side band for these emitters. On the other hand, the symmetry of the ZPL for group 2 is maintained at low temperature.

We now turn back to the range of ZPL energies observed within each group shown in Figure 1(b, c). It is well known that strain can induce shifts in the electronic structure and hence the optical transitions.[33] Such effects can be modeled using density functional theory (DFT) simulations, which we therefore used to calculate the optical response of a monolayer of hBN that contains the $N_BV_N$ defect which we described in detail in Ref 1,[1] and has similar photophysical properties to emitters in Group 2 of the present study. The DFT calculations (see Figure S7) show that the range of ZPL positions expected from 5% strain is comparable to that observed in each of the two emitter Groups shown in Figure 1. These results are in accord with strain induced exciton shifts that were observed for other 2D transition metal dichalcogenites, where strains of 3% resulted in 100 meV spectral shifts.[34, 35] Nonetheless, we acknowledge that the magnitude of strain needed to account for the entire observed range of ZPL positions within each group is high, and other effects such as variations in the dielectric environment of the emitters are likely to contribute to the observed shifts.

In conclusion, we have shown that room temperature multicolor single photon emission, with ZPLs spanning ~ 500 meV (200 nm), can be realized from defect centers in hBN multilayers. We present two robust methodologies to engineer the defects, based on annealing and electron beam irradiation. Moreover, we show that the emitters are stable even after annealing in harsh gaseous environments such as oxygen, hydrogen and ammonia. By analyzing spectral features of these emitters, we could infer that there are at least two groups of defects. Although the emitters in the two groups exhibit significant differences in their spectral characteristics, they share similar local phonon energies and therefore are likely to have similar chemical structure. Our work opens up new possibilities for employing quantum emitters in 2D materials for emerging applications in nanophotonics and nanoscale sensing devices.

## Methods

**Annealing hBN flakes under different gas environment**

Grid-marked native oxide Si (100) substrates were cleaned by ultrasonication in acetone and ethanol combined with light mechanical abrasion. Samples were prepared by drop-casting 100 μL of ethanol/water solution containing ~200 nm pristine h-BN flakes (Graphene Supermarket) onto marked substrates and allowed to dry.

Argon and $O_2$ annealing was carried out in a tube furnace (Lindberg Blue). The tube furnace was evacuated to low vacuum (~$10^{-3}$ Torr) by means of a scroll pump then purged for 30 minutes under 10 sccm of Ar or $O_2$ flow with pressure regulated at 1 Torr. The substrate was then heated under 10 sccm of argon flow or 500°C under 10 sccm of $O_2$ flow and held at a fixed temperature for 30 minutes, then allowed to cool to room temperature under continuous gas flow.

$H_2$ and $NH_3$ annealing was carried out in dedicated vacuum chamber. The chamber was evacuated to high vacuum (~$10^{-8}$ Torr) by means of a turbo molecular pump then purged for 30 minutes under 10 sccm of $H_2$ and $NH_3$ flow respectively with pressure regulated at 40 Torr. The substrate was then heated to 500°C under 10 sccm $H_2$ or $NH_3$ flow and held at this temperature for 30 minutes, then allowed to cool to room temperature under continuous gas flow.

**Electron-beam irradiation**

The substrates prepared for electron beam irradiation underwent photolithography procedures and subsequent metal deposition to create a hard mask grid. The substrates were then prepared as described above. The grid allowed for easy identification of areas to characterize optically before and after electron beam irradiation.

Electron-beam irradiation experiments were performed in a variable pressure FEI field emission gun scanning electron microscope. A low vacuum environment of $H_2O$ at a pressure of 8 Pa was used for all experiments. A focused beam was used in a raster scanning pattern to expose the hBN flakes as a function of time up to one hour over an area of 600 μm$^2$. An accelerating voltage of 15 kV and beam current of 1.4 nA were used for all electron beam irradiation experiments.

**Optical characterization**

A continuous wave (CW) 532 nm laser (Gem 532™, Laser Quantum Ltd.) was used for excitation and scanning. The laser was directed through a Glan-Taylor polarizer (Thorlabs Inc.) and a half waveplate, and focused onto the sample using a high numerical aperture (NA = 0.9, Nikon) objective lens. Scanning was performed either using an X-Y piezo scanning mirror (FSM-300™) or an X-Y-Z nanocube system (PI instruments). The collected light was filtered using a 532 nm dichroic mirror (532 nm laser BrightLine™, Semrock) and an additional long pass filter (Semrock). The signal was then coupled into a graded index fiber, where the fiber aperture serves as a confocal pinhole. A fiber splitter was used to direct the light to a spectrometer (Acton SpectraPro™, Princeton Instrument Inc.) or to two avalanche photodiodes (Excelitas Technologies™) used for single photon counting. Correlation measurements were done using a time-correlated single photon counting module (PicoHarp300™, PicoQuant™). The presented $g^2(\tau)$ curves were not corrected for background luminescence. Lifetime measurements were performed using a 510 nm pulsed laser excitation source (PiL051X™, Advanced Laser Diode Systems GmbH) with a 100 ps pulse width and a 20 MHz repetition rate. Low temperature PL spectroscopy was done at 14 K using a closed cycle refrigerating system cryostat (Janis CCS-XG-M/204N).

**Figures and captions**

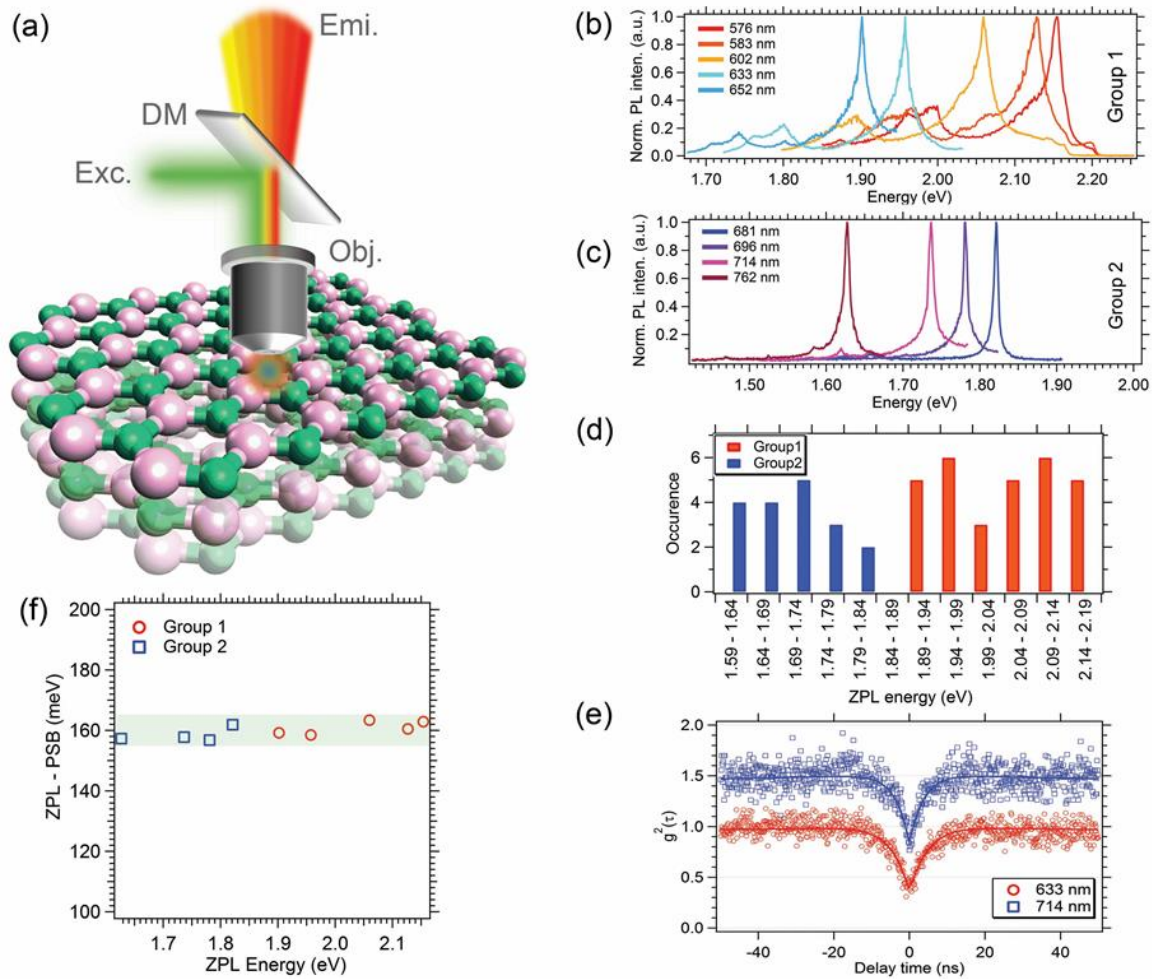

**Figure 1.** Multicolor photoluminescence from point defects in hBN. (a) Simplified schematic of the photoluminescence setup showing the excitation and emission of a defect center in a hBN lattice. The objective lens, dichroic mirror, excitation source and emission are denoted by Obj., DM, Exc., and Emi., respectively. (b) Five examples of emitters in Group 1 with ZPLs at 576 nm (2.15 eV), 583 nm (2.13 eV), 602 nm (2.06 eV), 633 nm (1.96 eV), and 652 nm (1.90 eV). (c) Four examples of emitters in Group 2 with ZPLs at 681 nm (1.82 eV), 696 nm (1.78 eV), 714 nm (1.74 eV), and 762 nm (1.63 eV). (d) Histogram of ZPL energy for numerous emitters in group 1 (red) and group 2 (blue). Each spectrum was acquired from a separate sample region at room temperature using a 300μW CW 532 nm laser. (e) Second order autocorrelation functions showing that $g^{(2)}(0) = 0.39$ and $0.34$, respectively. The $g^{(2)}(\tau)$ functions were acquired using an excitation power of 300 μW, an acquisition time of 20 sec, and were normalized (without background correction) and offset vertically for clarity. A neutral density filter was used to attenuate the signal generated by the 633 nm emitter. (f) Difference in the energy of the ZPL and PSB *versus* ZPL energy. The shaded band in (f) is a guide to the eye.

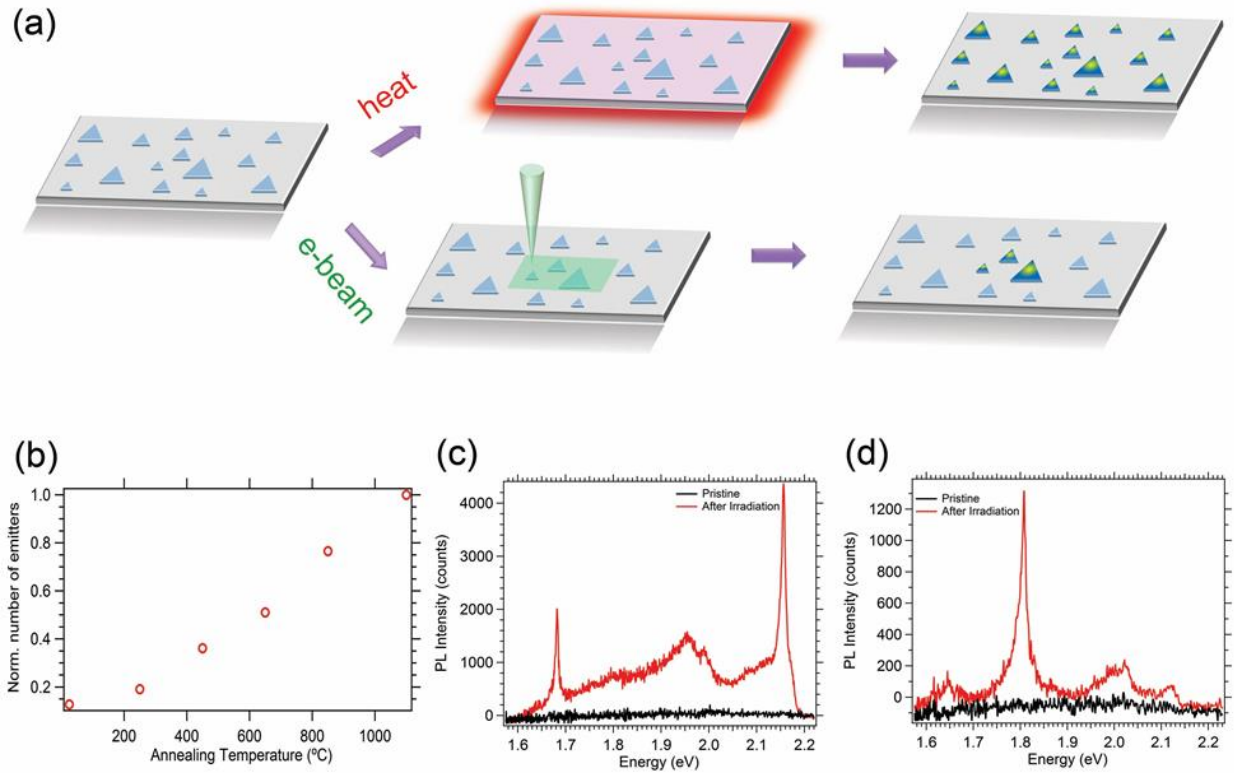

**Figure 2.** Generation of emitters in hBN. (a) Schematic illustration of two independent processes that yield emitters – annealing and electron beam irradiation. As-grown, dropcast hBN flakes are either annealed in an argon environment, or irradiated by an electron beam in a low vacuum $H_2O$ environment. (b) Normalized number of stable, bright single emitters as a function of annealing temperature found in hBN multilayers. Each data point was taken from a unique sample that was annealed at a singe temperature. (c, d) Examples of PL spectra from emitters fabricated by an electron beam. Each pair shows data recorded from a fixed sample region before (black curve) and after (red curve) electron irradiation.

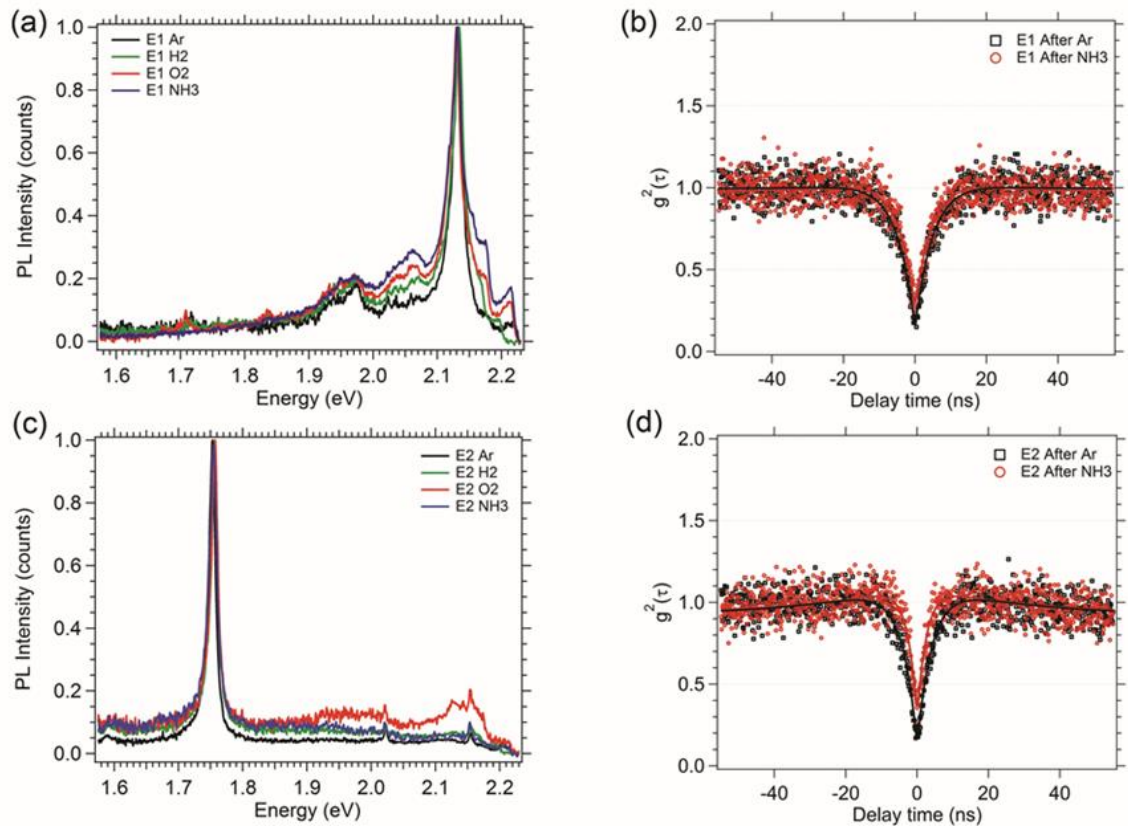

**Figure 3.** Stability of the emitters. (a,c) Normalized luminescence recorded at room temperature from emitters in Group 1 (E1) and Group 2 (E2) after sequential annealing in argon, hydrogen, oxygen and ammonia. (b, d) Corresponding antibunching measurements proving that the quantum nature of the defects persists after the sequential set of 30 min anneals performed in $H_2$, $O_2$ and $NH_3$ environments.

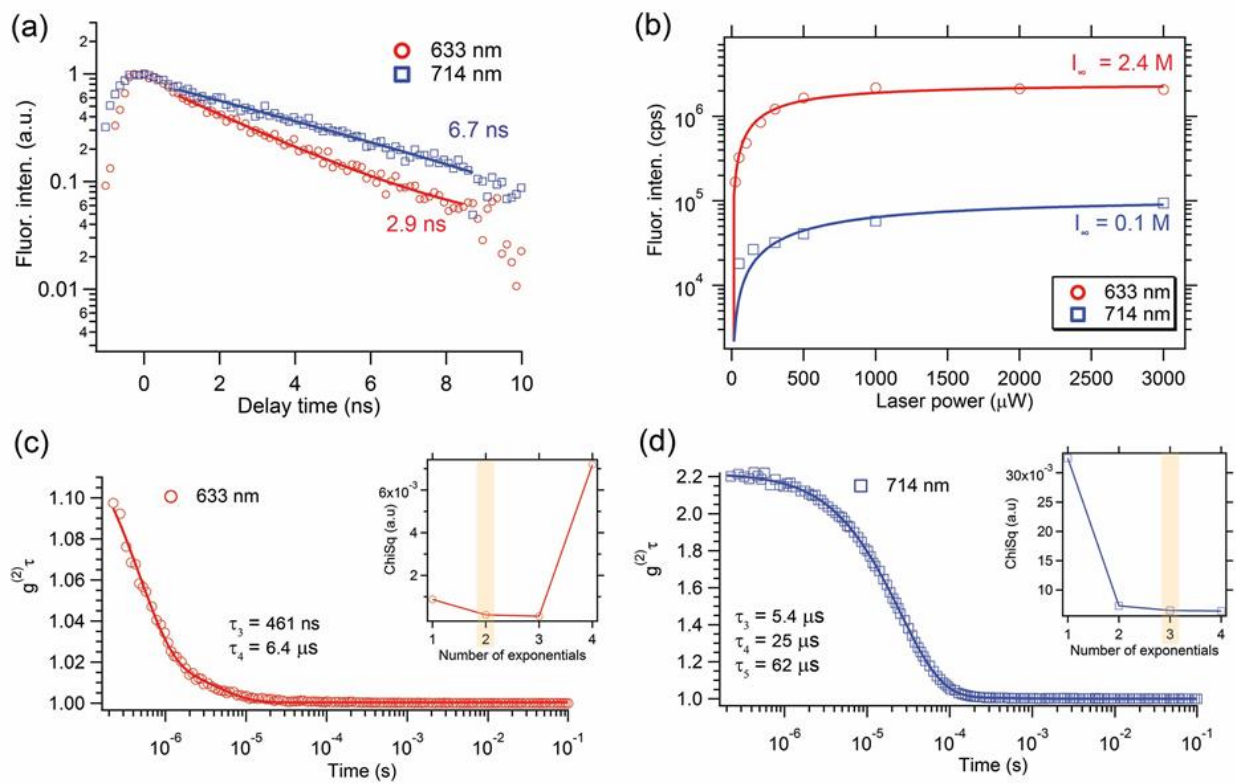

**Figure 4.** Photophysical properties of the defects. (a) Time-resolved fluorescence measurements showing radiative transition lifetimes of the emitters. A 80 μW, 510 nm pulsed laser with a repetition rate of 20 MHz and a pulse width of 100 ps was used as the excitation source. The solid lines are fits obtained using single exponential decay functions. (b) Fluorescence saturation curves and corresponding theoretical fits calculated using a three level model. (c, d) Second order autocorrelation function, $g^{(2)}(\tau)$, recorded over a longer time scale from the two color centers presented in Fig. 1 with ZPLs at 633 (c) nm and 714 nm (d). The corresponding solid traces are theoretical fits to the experimental data. Insets show the fitting residue $\chi^2$ *versus* the number of exponentials used in the fitting functions. The yellow bands indicate optimal fits realized when the number of exponentials and the residues are simultaneously minimized.

**Author Information**

*Corresponding authors: Milos.Toth@uts.edu.au; Igor.Aharonovich@uts.edu.au

*Notes: The authors declare no competing financial interest.

**Acknowledgements**

We thank Kerem Bray for technical help with the optical measurements and Fedor Jelezko for useful discussions. The work was supported in part by the Australian Research Council (Project Number DP140102721), FEI Company and by the Air Force Office of Scientific Research, United States Air Force. I. A. is the recipient of an Australian Research Council Discovery Early Career Research Award (Project Number DE130100592). G.G. acknowledges support by the Swiss National Science Foundation (SNSF). Measurements were supported in part by the Center for Excitonics, Center for Excitonics, an Energy Frontier Research Center funded by the U.S. Department of Energy, Office of Science, Office of Basic Energy Sciences under award no. DE-SC0001088. G.G. was supported in part by the National Science Foundation EFRI 2-DARE, Award Abstract #1542863.